\documentclass[a4,aps, showpacs,floatfix]{revtex4}
\usepackage{graphicx}
\usepackage{dcolumn}
\usepackage{color}
\usepackage{amsmath}
\usepackage{here}
\begin{document}

\title{Symmetry breaking in interacting ring-shaped superflows of ultacold
atoms}
\author{Artem Oliinyk$^{1}$, Igor Yatsuta$^{1}$, Boris Malomed$^2$,
Alexander Yakimenko$^{1}$}
\affiliation{$^1$ Department of Physics, Taras Shevchenko National University of Kyiv,
64/13, Volodymyrska Street, Kyiv 01601, Ukraine \\
$^2$ Department of Physical Electronics, Faculty of Engineering, and Center
for Light-Matter Interaction, Tel Aviv University, Tel Aviv 69978, Israel}

\begin{abstract}
We demonstrate that the evolution of superflows in interacting persistent
currents of ultracold gases is strongly affected by symmetry breaking of the
quantum vortex dynamics. We study counter-propagating superflows in a system
of two parallel rings in regimes of weak (a~Josephson junction with
tunneling through the barrier) and strong (rings merging across a reduced
barrier) interactions. For the weakly interacting toroidal Bose--Einstein
condensates, formation of rotational fluxons (Josephson vortices) is
associated with spontaneous breaking of the rotational symmetry of the
tunneling superflows. The influence of a controllable symmetry breaking on
the final state of~the~merging counter-propagating superflows is
investigated in the framework of a weakly dissipative mean-field model. It
is demonstrated that the population imbalance between the merging flows and
the breaking of the underlying rotational symmetry can drive the double-ring
system to final states with different angular momenta.
\end{abstract}

\maketitle

\section{Introduction}

Interacting Bose--Einstein condensates (BECs) suggest intriguing
possibilities for the investigation of spontaneous symmetry breaking in
quantum systems at the macroscopic level. In particular, coupled persistent
currents of ultracold atomic gases provide a possibility to investigate the
interaction of the~superflows in a tunable and controllable environment.
Using accessible experimental techniques, it is possible to consider a
variety of physical phenomena in this setting: from Josephson's effects in~the~regime of weak interactions to quantum Kelvin--Helmholtz instability for
merging rings.

Here we consider two parallel toroidal atomic BECs with opposite vorticities
in a three-dimensional (3D) trap. Previous theoretical investigations \cite{Brand10, Brand09, BrandPRL13, Brand18, Lesanovsky07, Luigi14,Davit13, Zhang13, Polo16, Haug18,PhysRevA.96.013620,Gallemi16, PhysRevA.97.053608,Tsubota2019, Tsubota2010_2, Tsubota2010} have drawn considerable interest to
systems of coupled circular BECs. In this vein, two parallel coaxial BEC
rings, separated in~the~axial direction by a potential barrier, were
considered in~the~context of the spontaneous generation of vortex lines~\cite{Montgomery10} and defects by means of the Kibble-Zurek mechanism~\cite{BrandPRL13}. It is worth to mention that binary systems with incoherent nonlinear
interaction between their components conserve the norms in each component
separately. Unlike the~present setting, such systems readily admit
stationary states with different vorticities and different chemical
potentials in the components. In particular, systems of this type give rise
to stable states with ``hidden vorticity'', i.e., ones with
opposite vorticities and equal norms in the two components, the
corresponding total angular moment being zero, as predicted in BEC~\cite{hidden1,hidden4,hidden6,hidden7,hidden8,hidden10,hidden11, hidden13} and
optics~\cite{hidden5,hidden9,hidden12}.

Tunneling of atoms through the potential barrier which separates two weakly
coupled ring-shaped condensates with different angular momenta leads to the
formation of Josephson vortices (rotational~fluxons) in the low-density area
between the rings~\cite{arxiv19}. This effect is characterized by zero total
tunneling flow between the rings. In the case of non-zero particle-number\
(population) imbalance between the rings, the Josephson vortices rotate and
bend. Splitting of Josephson vortices in the course of barrier reducing in
the system with weak dissipation makes it possible to reach states with
different final values of the total angular momenta of the merging rings,
depending on the initial population imbalance between them~\cite{arxiv19two}.

The main objective of the present work is investigation of the impact of
symmetries on interacting counter-propagating superflows in double-ring
systems. The analysis reveals three noteworthy effects: (i) The axial
symmetry of the superflows is spontaneously broken due to tunneling flow
across the~potential barrier and formation of Josephson vortices in the
low-density region between the rings. (ii)~When two axially-symmetric rings
with the counter-propagating superflows merge, the final state of the
toroidal condensate is never a ground state with zero angular momentum, as
might appear at the first sight. (iii) The ring-merging process and
topological charge of the final state can be controlled by the perturbation
of the trapping potential, specially adapted for the initiation of
symmetry-breaking of the system, and by tuning of the initial population
imbalance.

The rest of the paper is organized as follows. The model is formulated in~Section~\ref{sec2}. Results of~the~systematic analysis of the symmetry breaking in~double-ring systems are summarized in~Section~\ref{sec3}, separately for regimes of~weak and strong interactions. The paper is concluded by~Section~\ref{sec4}. 


\section{The model}
\label{sec2}

In modeling nonequilibrium phenomenology, such as quantum turbulence \cite%
{Tsubota13} or nucleation of vortices~\cite{PRA13}, dissipative effects are
of crucial importance for providing relaxation to equilibrium states. In
particular, the dissipation drives the drift of the vortex core to the edge
of the BEC cloud. Such~effects naturally arise in a trapped inhomogeneous
condensate due to its interaction with a~thermal component, and can be
captured phenomenologically by the dissipative GPE derived by Choi {\
et al}.~\cite{Choi,Proukakis}. Close to the thermodynamic equilibrium, the
weakly dissipative GPE is written as  
\begin{equation}
(i-\gamma )\hbar \frac{\partial \psi }{\partial t}=-\frac{\hbar ^{2}}{2M}
\nabla ^{2}\psi +V_{\text{ext}}(\mathbf{r},t)\psi +g|\psi |^{2}\psi -\mu
\psi ,  \label{GPE3D}
\end{equation}
where $g=4\pi a_{s}\hbar ^{2}/M$ is the nonlinearity strength, $M$ is the
atomic mass ($M=3.819\times 10^{-26}$ kg for $^{23}$Na atoms), $a_{s}$ is
the $s$-wave scattering length (positive $a_{s}=2.75$ nm, corresponding to
the~repulsion of sodium atoms, is used below), $\mu $ is the chemical
potential of the equilibrium state, and $\gamma \ll 1$ is~a~dimensionless
phenomenological dissipative parameter. This form of the  dissipative GPE
has been used extensively in previous studies of vortex  dynamics (see,
e.g.,~\cite{Tsubota13, PRA.77.023605, PRA.67.033610, PRA13}). It is based on the phenomenological approach, which correctly reproduces
experimental observations in~inhomogeneous BEC at finite temperature
provided that $T\ll T_c$. As is well known (see, e.g.~\cite{PhysRevA.81.023630}),
the~results of damped GPE for vortex dynamics are essentially the same as
produced by more general (and much more cumbersome) dissipative extensions
of GPE.  In what follows below, we assume $\gamma $ to be  spatially
uniform, and set $\gamma =0.03$, as in Refs.~\cite{Tsubota13, PRA13}. We
have verified that results reported below do not essentially  depend on a
specific value of $\gamma \ll 1$.

To elucidate the physical nature of the parameters $\gamma $ and $\mu $, it
is instructive to follow a simple derivation of the damped GPE (\ref{GPE3D})
presented in Ref.~\cite{Tsubota13}. It is assumed that the condensate,
described by wave function $\Psi $, exchanges energy and particles with a
thermal cloud (reservoir) of non-condensed atoms. The interaction with the
cloud is accounted for by an imaginary term on the right-hand side of~the~GPE:
\begin{equation}
i\hbar \frac{\partial \Psi }{\partial t}=\left( -\frac{\hbar ^{2}}{2M}\nabla
^{2}+V_{\text{ext}}(\mathbf{r},t)+g|\Psi |^{2}-i\Gamma \right) \Psi ,
\label{GPEDeriv}
\end{equation}%
which arises from the chemical-potential difference of the condensate and
thermal reservoir: \mbox{$\Gamma =\gamma (\mu -\mu _{\mathrm{th}})$}, where $\mu _{%
\mathrm{th}}$ is the chemical potential of the reservoir. Applying the
simplest approximation that $i\hbar \partial _{t}\Psi \simeq \mu \Psi $ and $%
\mu \simeq \mu _{\mathrm{th}}$, the gauge transformation of the wave
function, $\Psi =\psi e^{-i\mu t/\hbar}$ leads to dissipative GPE in the form of
Equation~(\ref{GPE3D}).

We consider a toroidal condensate, split by a narrow blue-detuned sheet beam
in upper and lower weakly coupled rings-shaped components. The respective
total trapping potential is  
\begin{equation}
V_{\mathrm{ext}}(\rho ,z,t)=\frac{1}{2}M\omega _{r}^{2}(\rho -\rho
_{0})^{2}+ \frac{1}{2}M\omega _{z}^{2}z^{2}+V_{\text{b}}(z,t),  \label{24}
\end{equation}
where $\rho \equiv \sqrt{x^{2}+y^{2}}$, and the sheet (barrier) potential is

\begin{equation}
V_{\text{b}}(z,t)=U_{\text{b}}(t)\exp \left( -\frac{1}{2}\frac{(\zeta
-z_{0})^{2}}{a^{2}}\right) ,  \label{Vb}
\end{equation}
with the time-dependent strength, 
\begin{equation}
U_{\text{b}}(t)=\left\{
\begin{array}{c}
(1-t/t_{d})u_{b},~\text{at~}~t<t_{d}, \\
0,~~\text{at~}~t>t_{d},%
\end{array}
\right.  \label{Ub}
\end{equation}

An experimentally relevant switching time is chosen as $t_{d}=0.015$ s, $
z_{0}$ being a possible shift of the barrier along the $z$-axis. The initial
barrier amplitude $u_{b}$ is well above the chemical potential $\mu $, so~that at $t=0$ two rings appear to be weakly coupled through the long
Josephson junction. To~address effects of the symmetry breaking on the
dynamics of the vortices in the course of~the~merger, in Equation~(\ref{Vb}) we
introduce, in  addition to time modulation (\ref{Ub}), uniform rotation of
the sheet beam  around the $x$-axis: 
\begin{equation}
\zeta =z\cos \left( {\Omega t}\right) -y\sin \left( {\Omega t}\right) ,
\label{rotation}
\end{equation}

Note that trapping potential (\ref{24})--(\ref{rotation}) with $\Omega =0$ (i.e., $\zeta =z$ in Equation~(\ref{rotation})) 
is symmetric with respect to rotation about the vertical ($z$) axis. Angular
velocity $\Omega $ in our simulations lies in the range from $\Omega _{0}=0$
(horizontal sheet beam) to $\Omega _{2}=2\pi \times 0.23$ Hz.  Thus the
final slope of the sheet beam with respect to the horizontal plane,  $\left(
x,y\right) $, at $t=t_{d}$ (when the barrier's amplitude vanishes,  as per
Equation~(\ref{Ub})), is small enough, to prevent full rotation of the  sheet
around the $x$-axis.

For numerical simulations of the 3D GPE we rescale time, $t\rightarrow
t\omega _{r}$, length, $\mathbf{r}\rightarrow \mathbf{r}/l_{r}$, the~chemical potential, $\mu \rightarrow \mu /(\hbar \omega _{r})$, the external
potential, $V_{\text{ext}}\rightarrow V_{\text{ext}}/(\hbar \omega _{r})$,
and the wave function, $\psi \rightarrow \psi \cdot l_{r}^{3/2}$, which
casts GPE (\ref{GPE3D}) in the following form:  
\begin{equation}
(i-\gamma )\frac{\partial \psi }{\partial t}=-\frac{1}{2}\nabla ^{2}\psi +{V}
_{\text{ext}}\psi -\mu \psi +g|\psi |^{2}\psi ,  \label{GPE_dimless}
\end{equation}
where the scaled positive nonlinearity strength is $g=4\pi a_{s}/l_{r}$, and
the scaled trapping potential~is  
\begin{equation}
V_{\text{ext}}=\frac{1}{2}(\rho -\rho _{0})^{2}+\frac{1}{2}A^{2}z^{2}+V_{
\text{b}},  \label{potential}
\end{equation}
with the aspect ratio of the toroidal trap,  
\begin{equation}
A=\omega _{z}/\omega _{r}~.  \label{A}
\end{equation}

It turns out that dynamics of quantum vortices, observed after the merger of
the rings, crucially depends on $A$~\cite{arxiv19two}, while the variation
of other parameters of condensate does not change our main conclusions
qualitatively. In this work we  concentrate on the pancake-shaped trapping
potential with typical values of  the trapping frequencies~\cite{Wright13,FredPRL14}: $\omega _{r}=2\pi \times  123$ Hz and $\omega
_{z}=2\pi \times 600$ Hz, hence $A=4.88$, the oscillator  length of the
radial trapping potential is $l_{r}=\sqrt{\hbar /\left( M\omega _{r}\right) }%
=1.84$ $\mu$m, 
$\rho _{0}=19.23$ $\mu$m, and $%
g=1.88\times 10^{-2}$. Scaled parameters of the potentials in Equations~( \ref{Vb}%
) and (\ref{Ub}) are fixed to be $a=0.3,u_{b}=80$, which make it  possible
to produce generic results. 
Below, we use the same notation for the scaled wave function $\psi $,
spatial coordinates $\left( x,y,z\right) $, and time $t$ as above, as it
will produce no confusion.


\section{Symmetry breaking in coupled condensate rings}

\label{sec3}

\subsection{Spontaneous Symmetry Breaking in a Stationary Hybrid Vortex
Structure}

First, we use the imaginary-time-propagation (ITP)\ numerical method to
obtain a steady-state solution of Equation~(\ref{GPE_dimless}) with $\gamma =0$.
``Hybrid'' states, with coupled rings carrying different
vorticities~\cite{NJP14}, are produced by this method, starting from the
following initial state:  
\begin{equation}
\Psi (\mathbf{r})=|\Psi _{0}(x,y,z)|e^{iS(z)\theta },  \label{input}
\end{equation}
where $\theta $ is the azimuthal  angle, $S(z)=m_{1}$ for $z<z_{0}$  and $%
S(z)=m_{2}$ for $z\geq z_{0}$. The ITP converges to steady states with
required accuracy for an arbitrary input $\Psi _{0}(x,y,z)$ in Equation~(\ref%
{input}) with a~fixed~norm:  
\begin{equation}
\langle \Psi _{0}|\Psi _{0}\rangle =N\equiv N_{1}+N_{2},  \label{N}
\end{equation}
where $N_{1}$ and $N_{2}$ are scaled populations in the bottom and top
rings, respectively:  
\begin{equation}
N_{1,2}=\int_{V_{1,2}}|\Psi _{0}(\mathbf{r})|^{2}d\mathbf{r},  \label{intN}
\end{equation}
with integration areas $V_{1,2}$ corresponding to half-space $z<z_{0}$ for
lower ring, and $z\geq z_{0}$ for the upper one, respectively.

We note that, by shifting center $z_{0}$ of the splitting barrier (\ref{Vb}), it is easy to prepare an initial state with a dominant population in the
ring with topological charge $m_{1}$ ($N_{1}>N_{2}$ for $z_{0}>0$) or $m_{2}$
($N_{2}>N_{1}$ for $z_{0}<0$), the respective asymmetry parameter being

\begin{equation}
P=(N_{1}-N_{2})/(N_{1}+N_{2}).  \label{P}
\end{equation}

\begin{figure}[H]
\centering
\includegraphics[width=6.8in]{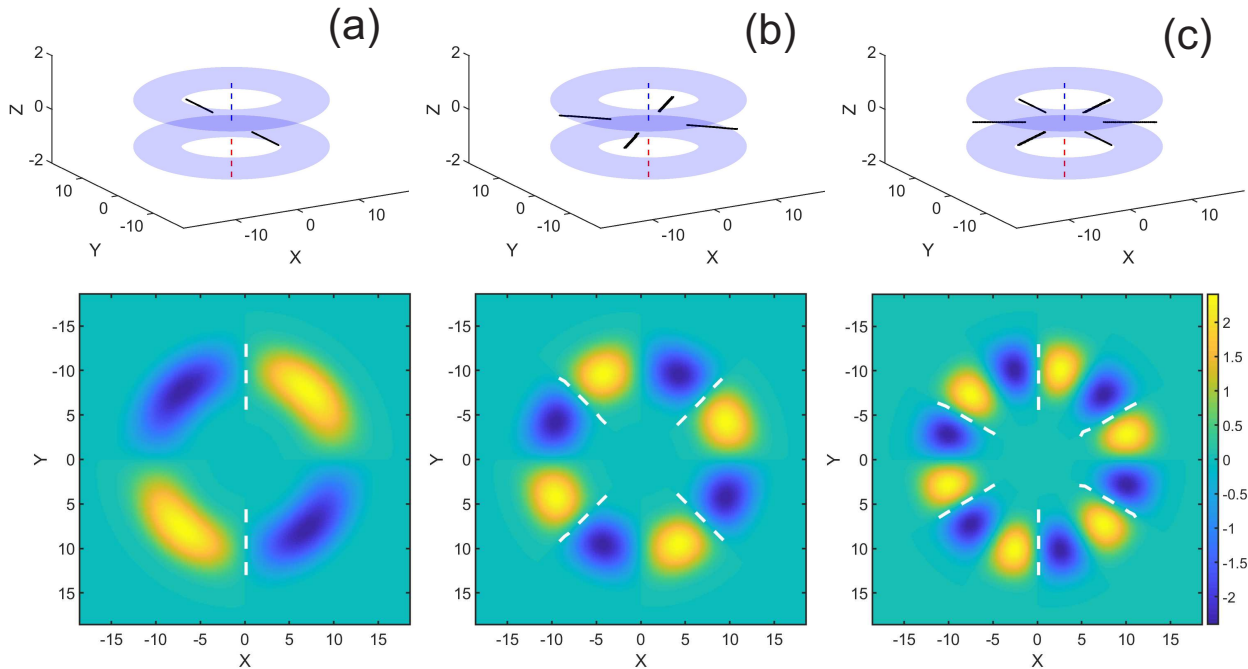}
\caption{(Color online) Hybrid vortex stationary states with hidden
vorticity and zero population imbalance, $P=0$ (see Equation~(\protect\ref{P})):
(\textbf{a}) $\left( m_{1,}m_{2}\right) =(+1,-1)$, (\textbf{b}) $(+2,-2)$, (\textbf{c}) $(+3,-3)$.
Shown~are density isosurfaces (the top row) and the $z$-component of the
corresponding tunnel-flow density distribution through the barrier, $%
j_{z}(x,y,z=0)$ (the bottom row). The cores of the Josephson vortices are
indicated by black lines in the top row, and by white dashed lines in the
bottom one. Vertical red and blue dashed lines designate cores of the
counter-propagating persistent currents in the two rings.}
\label{123cores} 
\end{figure}

Here we consider stationary hybrid states with hidden vorticities: $
m_{1}=-m_{2}$. Figure \ref{123cores} shows three different states produced
by the ITP method for $m_{1}$ $=1,2,3$ and $P=0$ with high accuracy.
Different directions of the flow in the top and bottom rings inevitably lead
to the appearance of Josephson vortices (fluxons), with the number of fluxon
cores $N_{J}=|m_{1}-m_{2}|$, as pointed out in our previous work \cite%
{arxiv19two}. To produce additional description of the vortices, we need to
consider the flow in~both parts of the condensate.

The density of the superflow in scaled units is defined by the usual
expression:  
\begin{equation}
\mathbf{j}(\mathbf{r},t)=-\frac{i}{2}\left\{ \psi ^{\ast }(\mathbf{r},t){\
\nabla }\psi (\mathbf{r},t)-\psi (\mathbf{r},t){\nabla }\psi ^{\ast }(
\mathbf{r},t)\right\} .  \label{FlowDensDef}
\end{equation}

As it was shown in~\cite{arxiv19}, in the case of BEC loaded in
parallel-coupled ring-shaped traps, azimuthal distribution of the $j_{z}$
component of the flow can be accurately described in the Galerkin
(finite-mode) approximation. This approach yields the following azimuthal
distribution for the flow tunneling across the potential barrier:  
\begin{equation}
j_{z}\sim \sin [\omega t+(m_{2}-m_{1})\theta +\Delta ],  \label{GalFlow}
\end{equation}
where the scaled angular velocity of the fluxon, 
\begin{equation}
\omega =\mu _{1}-\mu _{2},  \label{omega}
\end{equation}
is defined by chemical potential difference, and $\Delta$ is a constant
phase difference between the upper and  lower rings.

The position of fluxon cores is determined by the flow distribution, given
by Equation~(\ref{FlowDensDef}), and~rotation of this distribution  around the $z$%
-axis will cause a similar rotation around the same axis of  the
radially-oriented Josephson vortices without changing their relative
position. The initial vertical position of the fluxons is imposed by $z_{0}$
, meanwhile their angular positions may be arbitrary, being defined by the
value of $\Delta $. It is remarkable that the spatial structure of the
tunneling flows and formation of the Josephson vortices \emph{{spontaneously
breaks the azimuthal symmetry}} 
of the stationary hybrid vortex states with
hidden vorticity, even in a fully symmetric trapping potential, cf. Ref.
\cite{NJP14}. It also follows from Equation~(\ref{GalFlow}) that, in the case of
the population imbalance ($P\neq 0$ with nonzero chemical potential
difference, and, accordingly, $\omega \neq 0$, see Equation~(\ref{omega})),
vortices perform rotational motion with angular velocity $\omega $. Here and
in the next section we consider the case of $\Delta =0$. The~corresponding
flow distribution and positions of the vortices for $P=0$ are shown in
Figure  \ref{123cores}. Further we consider evolution of the stationary
hybrid  states with a reduced barrier for different values of imbalance $P$~in~the~range from $-1$ to $1$.

We stress that the outlined mechanism of the rotational symmetry breaking in
atomic BECs is not related directly to the spontaneous symmetry breaking in
the quantum field theory, or BCS theory of superconductivity. However, in a
certain sense, Josephson vortices play a similar role in the present setting
as the Nambu-Goldstone bosons in the above-mentioned contexts \cite%
{Arraut19,Nambu04, Brauner10,NIELSEN1976445}. In particular, the~dispersion
law of the vortices is quadratic, for a straightforward reason: if the
vortex is slowly moving, its kinetic energy is proportional to the square of
the velocity.

It is noteworthy that the hybrid stationary states, which also play the  role
of the initial conditions for dynamical simulations, feature the  symmetry
under the rotation around the $z$-axis by an angle of $2\pi k/N_{J}$
(discrete rotational symmetry), where $k$ is integer and $%
N_{J}=|m_{1}-m_{2}|  $ is the even number of radially-oriented Josephson
vortices located in the  junction between the counter-propagating
superflows. In the next section, we  investigate the influence of this
symmetry on the evolution of the~merging  persistent currents.


\subsection{Influence of the symmetry on dynamics of the merging rings.}

We concentrate here on simulations of dynamics of the merging rings with
single-charged counter-propagating persistent currents, $
(m_{1},m_{2})=(+1,-1)$, in a pancake-shaped toroidal trap. Note~that a
similar double-ring system, but loaded in a trap elongated in\ the $z$-direction, was investigated in~\cite{arxiv19two}. As pointed out in that
work, vortex dynamics and\ the relaxation process are strongly affected by
the value of aspect ratio $A$, see Equation~(\ref{A}). Our analysis in the
present work demonstrates that the evolution of the merging rings is
substantially affected by the symmetry of~the~pancake-shaped trap. Detailed
analysis of the impact of the symmetry breaking on the evolution of the
merging persistent currents is the main objective of this subsection.

We used the split-step fast-Fourier-transform method for numerical
simulations of the dissipative GPE (\ref{GPE_dimless}) in real time. The
total number of atoms (\ref{intN}), $z$-component $L_{z}$ of the angular
momentum  
\begin{equation}
\mathbf{L}=-\frac{i}{2}\int \left\{ \psi ^{\ast }\left[ \mathbf{r}\times
\nabla \psi \right] -\psi \left[ \mathbf{r}\times \nabla \psi ^{\ast }\right]
\right\} d\mathbf{r},  \label{L}
\end{equation}
and energy  
\begin{equation}
E=\int \left[ \frac{1}{2}|\nabla \psi |^{2}+V_{\text{ext}}(\mathbf{r})|\psi
|^{2}+\frac{g}{2}|\psi |^{4}\right] d\mathbf{r},  \label{E}
\end{equation}
are not conserved in the dissipative setting described by GPE (\ref%
{GPE_dimless}) with $\gamma \neq 0$. For the time-independent trapping
potential, the wave function evolves towards the steady state corresponding
to the chemical potential $\mu $. However, for the time-dependent external
potential and constant chemical potential $\mu $, the temporal evolution of
the number of atoms is inconsistent with typical experimental observations,
where the number of particles exponentially reduces with time: $N(t)=N(0)$%
exp $(-t/t_{0})$, where $t_{0}=10$ s is a characteristic lifetime of the BEC
in  the experiment~\cite{Hadzibabic}. To reproduce correctly the time
evolution  of the number of atoms we use a method similar to the approach
suggested in  Ref.~\cite{Tsubota13} for quantum hydrodynamics. Namely, an
appropriate  decay rate of $N$ is forced by adjustment of chemical potential
$\mu (t)$ of  the equilibrium state at each time step in the course of the
simulations.  The initial total number of atoms is $N(0)=6\times 10^{5}$.

\begin{figure}[H]
\centering
\includegraphics[width=6.8in]{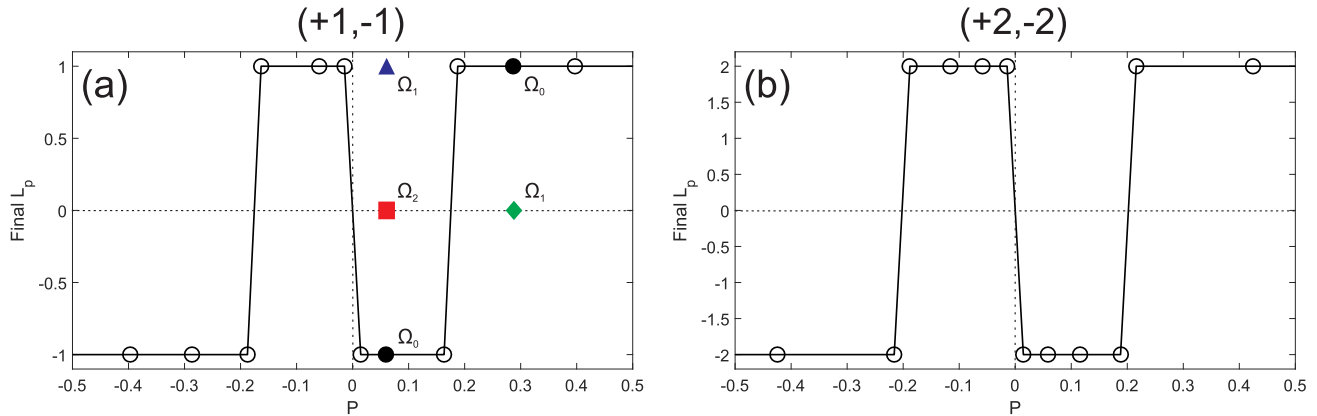}
\caption{(Color online) The final value of the total angular momentum per
particle, \mbox{$L_{p}=L_{z}/N$}, for the merging rings with initial vorticities $%
(m_{1},m_{2})$, as a function of initial imbalance $P$: \mbox{(\textbf{a})~$(m_{1}=+1,m_{2}=-1)$,} (\textbf{b}) $(m_{1}=+2,m_{2}=-2)$. Solid black lines with
circles represent the final states for the axially symmetric trapping
potential (the horizontal sheet beam, which corresponds to $\Omega =0$ in
Equation~(\protect\ref{rotation})). Surprisingly, merging counter-rotating flows
in the axially-symmetric trap never evolve towards the non-rotating ground
state, with $L_{p}=0$, even for small imbalances, $|P|\ll 1$. The vorticity
of the final state is imposed by the \emph{less populated} component if $%
|P|<P_{\text{cr}}\approx 0.1755$ (see~Equation~(\protect\ref{Pcr})) for initial
vorticities $(+1,-1)$, and $P_{\text{cr}}\approx 0.21$ for $(+2,-2)$ (this
counter-intuitive result is explained in the main text), and by the stronger
component if $|P|>P_{\text{cr}}$. The impact of the~symmetry breaking of the
trapping potential is illustrated for the setup with initial vorticities $%
(+1,-1)$ by examples of the final states for two values of the imbalance, $%
P=0.06$ and $P=0.29$. For~$P=0.06$, the filled black circle corresponds to
the nonrotating (horizontal) barrier (with $\Omega _{0}=0$ in~Equation~(\protect
\ref{rotation}), see Figure \protect\ref{angless}), while the blue triangle
and red square correspond to the barrier rotating around the $x$-axis
barrier with angular velocity $\Omega _{1}=2\protect\pi \times 0.11$ Hz (see
Figure \protect\ref{ang}) and $\Omega _{2}=2\protect\pi \times 0.23$~Hz (see
Figure \protect\ref{bigang}), 
respectively. For $P=0.29$, the filled black
circle corresponds to the $\Omega _{0}=0$, and the~green diamond corresponds
to $\Omega _{1}=2\protect\pi \times 0.11$ Hz. Note that, above the threshold
imbalance, $P>P_{\text{cr}}$, the~final state with $L_{p}=-1$ is never
observed even for the system with broken symmetry. }
\label{1m1Lp}
\end{figure}

Figure \ref{1m1Lp}a summarizes our findings. It shows, by means of black
lines with circles, the final value of the angular momentum per particle, $
L_{p}=L_{z}/N$, as a function of population imbalance $P$. The~integer
values of $L_{p}$ are obtained after completion of the merger of the rings
separated by the~horizontal sheet beam ($\Omega =0$ in Equation~(\ref{rotation}))
and subsequent relaxation process towards a stationary persistent current in
the toroidal trap. Two remarkable features are seen: (i) The final
non-rotating ground state, with $L_{p}=0$, \emph{{is not produced}} by the
simulations even for practically symmetric states with $P\approx 0$ (so~that
the initial total angular momentum is close to zero). Instead, the merging
rings evolve into an overall-vortex ($m=+1$) or antivortex ($m=-1$) state.
(ii) For the initial states with the imbalance below a well-defined critical
value,  
\begin{equation}
|P|<P_{\text{cr}}\approx 0.1755,  \label{Pcr}
\end{equation}
the final angular momentum is, counterintuitively, determined by the
less populated ring, while for $|P|>P_{\text{cr}}$ the more populated ring
imposes its angular-momentum state onto the whole condensate. This
unexpected conclusion is explained below.

We have performed similar analysis for double-charged counter-propagating
persistent currents, with $(m_{1},m_{2})=(+2,-2)$. As is seen from Figure %
\ref{1m1Lp}b, final states solely with $L_{p}=+2$ and $L_{p}=-2$ are
observed  for the symmetric trapping potential with the horizontal sheet
beam. The  critical value of the imbalance appears to be $P_{\text{cr}%
}\approx 0.21$.  


Note that the merging rings with vorticities $(2,0)$, which we investigated
in Ref.~\cite{arxiv19two}, also evolve into the final state with the total
angular momentum $m=0$ or $m=2$, while the single-charged state, with $m=1$,
was not realized even when the angular momentum per particle of the initial
state was $L_{p}=1$. A more populated component with $m_{1}=2$ imposes its
angular momentum onto the final state, provided that the initial imbalance
takes values above some critical value $P_{\text{cr}}$. However, the~values
of $P_{\text{cr}}$ found in~\cite{arxiv19two} for the $(2,0)$ set are
different from those $\pm P_{\text{cr}}$ for the set of the merging rings
with $\left( m_{1},m_{2}\right) =(+1,-1)$, given by Equation~(\ref{Pcr}). These
differences are not surprising, as the set of $(+1,-1)$ features an obvious
symmetry with respect to the two components.

\begin{figure}[H]
\centering
\includegraphics[width=6.8in]{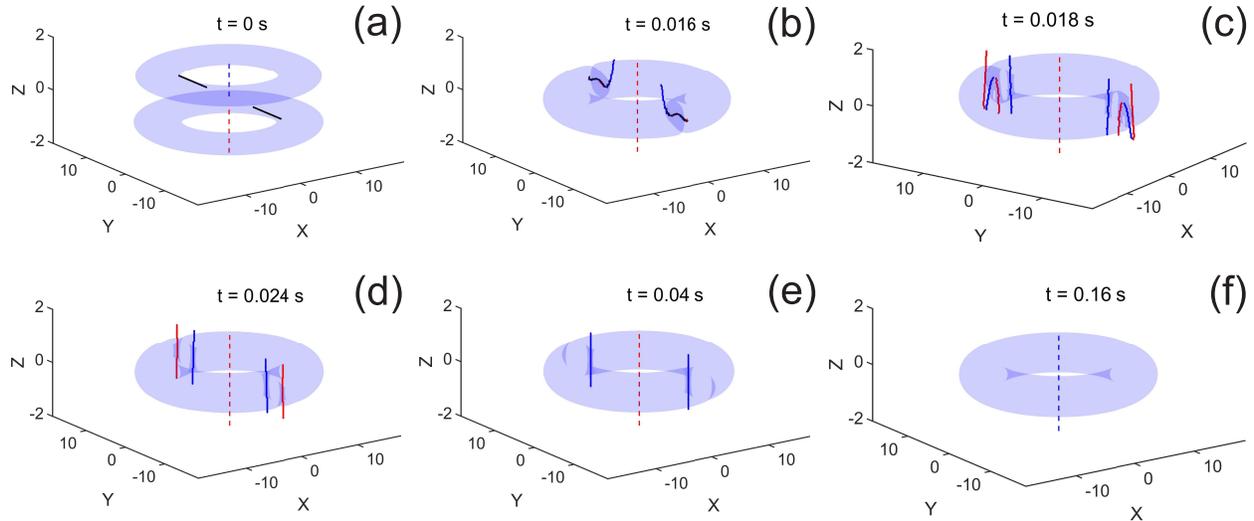}
\caption{(Color online) The evolution of the merging rings without symmetry
breaking ($\Omega =0$ in~Equation~(\protect\ref{rotation})). The barrier
separating two rings is switched off at $t>t_{d}=0.015$ s. Red (blue) lines
indicates positions of the vortex (antivortex) core. The population of the
bottom ring, with $m_{1}=+1$, is~slightly larger than in the top one, with $%
m_{2}=-1$ (imbalance parameter (\protect\ref{P}) is $P=0.06$). The~final
state has $m=-1$, as shown in Figure \protect\ref{1m1Lp}a by the filled
black circle. Note that,in the course of the~evolution of the~merging
counter-rotating flows in the axially symmetric trap, the discrete
rotational symmetry is observed for the positions of the vortex cores with
respect to the rotation around the $z$-axis by an~angle of $\protect\pi $.
The symmetric drift of two diametrically opposite antivortices towards the
central hole leads to subsequent annihilation of the central vortex and
relaxation of the toroidal condensate into a~final antivortex state.}
\label{angless}

\end{figure}

Typical vortex dynamics for the system with the horizontal sheet beam
(without the discrete rotational symmetry breaking, i.e., with $\Omega =0$
in~Equation~(\ref{rotation})), for imbalance parameter $P=0.06$, is~shown in
Figure  \ref{angless}. The initial pair of the Josephson vortices are
bending in the  course of the~merger of the rings. The horizontal
orientation of the  Josephson vortices is not energetically preferable in~the~pancake-shaped  setup: vortex lines tend to change the orientation and
to be directed along  the~$z$-axis (see also~\cite{arxiv19two}). As is seen
in Figure \ref{angless}b, the initial central antivortex (shown by the
dashed blue line in the  upper, less populated, ring in Figure \ref{angless}a) splits into two  antivortex lines (blue curves at the end of the
Josephson vortices) and the  vortex one (the red dashed line) immediately
when the~separating barrier  vanishes. This process allows the central
vortex, which is initially located only in~the~bottom, more populated, ring
to occupy also the top, less populated, area of  the eventual state. Figure %
\ref{angless}d shows the state when the  splitting and reorientation of
the initial Josephson vortices is completed.  Vertically oriented vortices
and antivortices move from high-density to a  low-density area due to the
dissipation, hence the system relaxes to the  available state with the lower
energy. This~process is illustrated in Figure  \ref{angless}d--f: both
vortices (red lines) shown in Figure \ref{angless}d  disappear at the~periphery of the condensate, while both antivortices (blue  lines) move to
the inner edge of the~toroidal condensate, and one of these  antivortex
annihilates with the central initial vortex (the~red dashed  line). Thus,
the final state is determined by the less populated subsystem  carrying the
antivortex (see~Figure~\ref{angless}f).

As said above, this result seems counter-intuitive, because in the initial
state vorticity $m_{1}=+1$ of the bottom ring was more populated than its
top-ring counterpart with $m_{2}=-1$. Nevertheless, it is reasonable to
expect that the subsystem with $m_{1}=+1$ imposes its vorticity onto the
final state when the population in the bottom ring substantially dominates
(at least in the limit of $P\rightarrow 1$). Indeed, at~values of the
initial imbalance exceeding the threshold value given by Equation~(\ref{Pcr}),
the final state is the single-charged persistent current with $m=+1$ {[}see
Figure \ref{1m1Lp}a{]}. 

Analysis of the symmetry in the vortex dynamics provides insight into
unusual properties of the~merging persistent currents. The dynamics of the
vortices maintains the discrete rotational symmetry imposed by the symmetry
of the external trapping potential. In particular, it is seen in~Figure~\ref%
{angless} that the dynamics of the vortex cores obeys the symmetry under the
rotation around the $z$-axis by $\pi $, for both groups of vortices, which
appear after decay of two Josephson vortices [see~Figure~\ref{angless}b--e]. Thus, in addition to the vortex line originated from the stronger
component, either none or even number of vortices (antivortices) can be
simultaneously trapped in the central hole of the~toroidal condensate after
the merger of the double-ring set with counter-rotating flows $m_{1}=-m_{2}$. For example, at $P<P_{\text{cr}}$ the symmetric drift of two diametrically
opposite antivortices towards the central hole (as seen in Figure \ref%
{angless}e) leads to subsequent annihilation of the~central vortex, and the~relaxation of the toroidal condensate into the final antivortex state
with $ m=-1$ (see Figure \ref{angless}f). At $P>P_{\text{cr}}$, after the
merger  of the rings, the split vortices and antivortices (similar to those
shown in~Figure \ref{angless}d) are located close to the external edge
of the  condensate, in comparison to the case of $P<P_{\text{cr}}$
considered above.  Thus all these vortices tend to decay at the external
periphery, and the  final state is determined by the vorticity of the
initially more populated  ring with $m_{1}=+1$.

This is why the final topological charge $L_{p}$ appears to be imposed by
the stronger component (with~$m_{1}=+1$) at $P>P_{\text{cr}}$, and by weaker
one, with $m_{2}=-1$, at $P>P_{\text{cr}}$, while the~ground state with $%
L_{p}=0$ is forbidden as an outcome of the relaxation of the merging rings
in~the~axially-symmetric trapping potential. These properties may be
expected for any setting with two counter-rotating superflows and an even
number of Josephson vortices in the initial state, as illustrated for
double-charged persistent currents in~Figure~\ref{1m1Lp}b. When the
superlflows merge, the symmetric dynamics of the relaxation process,
outlined above, excludes the final state $L_{p}=0$, and explains why, for $%
P<P_{\text{cr}}$, the final angular momentum of the merged system is
determined by the less populated ring. Note that these properties are not
affected by the dissipative effects, since they are completely determined by
the~symmetry of the system.

It is well known that the toroidal condensate carrying a persistent current
is a remarkable example of quantum multistable systems with local minima of
the energy for a given integer value of the~per-particle angular momentum $
L_{p}$ (see, e.g.~\cite{RRP2015}). Thus, as the result of the merger of the~double-ring set, the final states with different topological charges can be
realized. However, the above results also demonstrate that some values of
the angular momentum (vorticity) cannot be produced by the merger in the
symmetric system. This raises the question as to whether the final states
inaccessible for the symmetric system can be observed if the rotational
symmetry of the trapping potential is broken.

Certainly, there are many ways to break the symmetry of the trapping
potential. In this work we~break the symmetry of the $(+1,-1)$ setup in a
controllable way by the slow rotation of the barrier around the horizontal
axis in the course of the merger, see Equation~(\ref{rotation}). The rotation is
a simple perturbation, which can be easily realized experimentally. The
breaking of the azimuthal symmetry of the potential, induced by the
rotation, leads to redistribution of the condensate density in~the~course of
the merger, which strongly affects subsequent dynamics of the vortices, when
the barrier is switched~off.



Obviously, an atomic cloud is never purely symmetric in a real experiment,
due to imperfections of the trapping potential and irregular time-dependent
fluctuations of the system's parameters. Then,~the~question arises if
unusual manifestations of the symmetry of the merging persistent currents,
predicted above, are experimentally observable, or maybe any non-negligible
symmetry-breaking perturbation essentially affects the final vorticity
state. To address this issue, we have performed extensive simulations,
including stronger or weaker symmetry-breaking perturbations. It turns out
that, when the sheet beam rotates very slowly ($\Omega /(2\pi )\ll 0.1$ Hz
in Equation~(\ref{rotation})), the symmetry breaking does not produce any visible
effect on the dynamics of the vortices, with the vorticity of the~final
state being the same as for the horizontal sheet beam (with $\Omega =0$).
The impact of~the~symmetry breaking becomes essential as the angular
velocity of the rotating sheet beam increases.

\begin{figure}[H]
\centering
\includegraphics[width=6.8in]{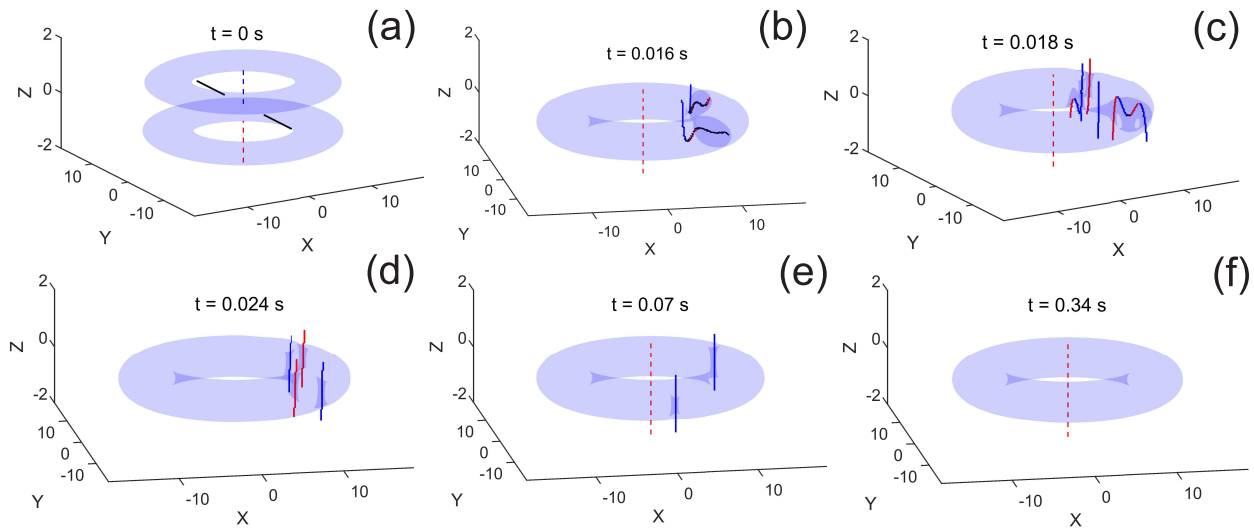}
\caption{(Color online) The evolution of the merging rings affected by
symmetry breaking which is induced by slow rotation of the sheet beam (\protect\ref{rotation}) around the $x$-axis, with angular velocity \mbox{$\Omega
_{1}=2\protect\pi \times 0.11$ Hz.} Note that the system with the broken
symmetry evolves towards the final topological charge (vorticity) $m=+1$,
while, for the same value of imbalance (\protect\ref{P}), $P=0.06$ (the~population of the~bottom ring with $m_{1}=+1$ slightly dominates over the~top one, with $m_{2}=-1$), the final state of the~axially-symmetric system
has $m=-1$ (see Figure \protect\ref{angless}).} 
\label{ang}
\end{figure}

Typical examples of the relaxation dynamics for small imbalance $P=0.06$ and
broken symmetry are shown in Figure \ref{ang} for $\Omega _{1}=2\pi \times
0.11  $ Hz (the final topological charge is $m=+1$) and in~Figure~\ref%
{bigang} for $ \Omega _{2}=2\pi \times 0.23$ Hz (the system relaxes to the
non-rotating  ground state, with $m=0$). Note that, for the same imbalance
but without the  rotation of the barrier ($\Omega =0$), the final state is $%
m=-1$ (see~Figure~\ref{1m1Lp}a, (\ref{angless})).

\begin{figure}[H]
\centering
\includegraphics[width=6.8in]{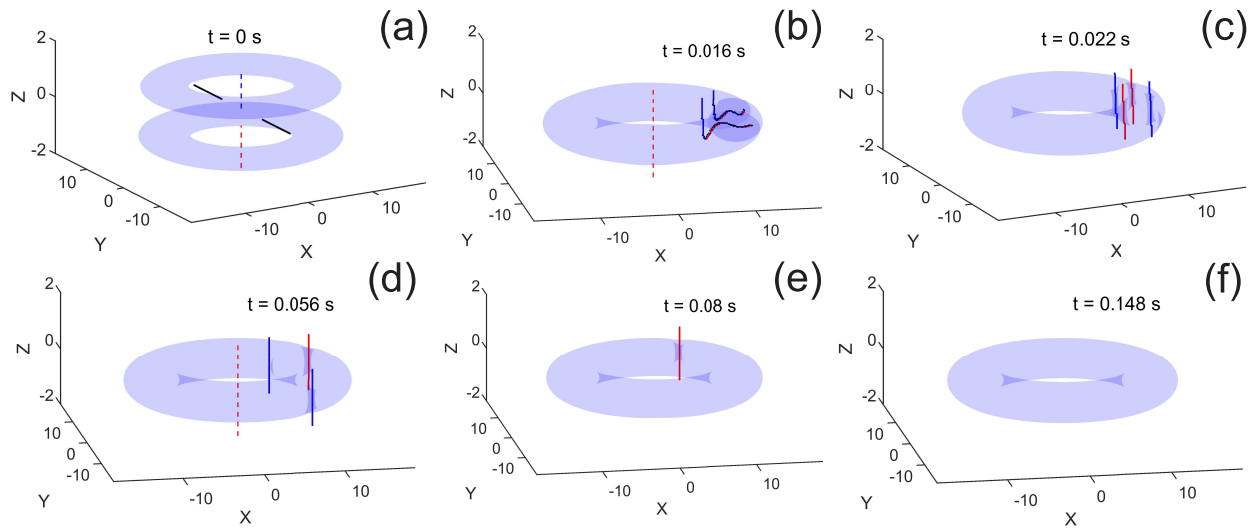}
\caption{(Color online) The evolution of the merging rings affected by
stronger, than in the case of Figure \protect\ref{ang}, symmetry breaking,
imposed by rotation (\protect\ref{rotation}) with angular velocity $\Omega
_{2}=2\protect\pi \times 0.23$ Hz. The~population in the bottom ring with $%
m_{1}=+1$ slightly dominates over the top one, with $m_{2}=-1$ (the
respective imbalance parameter (\protect\ref{P}) is $P=0.06$, as well as in
Figures~\ref{angless}~and~\ref{ang}). Being controlled by the
barrier's angular velocity, $\Omega $, the symmetry breaking drives the
merging rings to final states with different topological charges
(vorticities). In this case, when the axial symmetry is \emph{strongly}
broken, the final \textit{nonrotating} state is established, with vorticity $%
m=0$. Surprisingly, merging counter-rotating persistent currents evolve into
a nonrotating final state neither for the symmetric system (see~Figure~\ref{angless}, where $\Omega _{0}=0$ and the final topological
charge is $m=-1$), nor for a weakly asymmetric trapping potential (see
Figure \protect\ref{ang}, where $\Omega _{1}=2\protect\pi \times 0.11$ Hz
and the final topological charge is $m=+1$).} 
\label{bigang}
\end{figure}

Remarkably, when the barrier is switched off, the positions of the split
vortices are not diametrically opposite, as it was observed in the absence
of the symmetry breaking ($\Omega =0$). In~the~course of the merger, if $
\Omega \neq 0$, the initial Josephson vortices move toward each other until
they break, bend and split (see Figures~\ref{ang}b,c~and~\ref
{bigang}b,c).

Recall that the initial hybrid state has different vorticities, $m_{1}=+1$
and $m_{2}=-1$, in their two axially separated parts (see Figure \ref{ang}a). Both vorticities share a common vertically oriented axis threading the
separated double-ring system, and $|m_{1}-m_{2}|$ radially oriented
rotational fluxons inside the separating barrier. The remarkable moment of
the evolution of the merging persistent currents is illustrated in Figure %
\ref{ang}b: as soon as the barrier vanishes, the central vortex line (the~dashed red line in Figure \ref{ang}a),\ which originates from the
dominating bottom ring, still exists at $t>t_{d}$ (see~Figure~\ref{ang}b),  while the antivortex from the upper ring (the dashed blue line in
Figure \ref{ang}a) splits into one vortex and two antivortices, which are
shown by  solid blue lines attached to the bending Josephson vortices in~Figure \ref{ang}b. As the result, the central hole contains a single
vortex line (the  dashed red line in~Figure~\ref{ang}b) threading the
whole toroidal  condensate. Thus, similar transformations are observed for
the axisymmetric  trap (Figure \ref{angless}b) and in the case of broken
symmetry (Figures \ref{ang}b~and~\ref{bigang}b). However, for the system
with the horizontal  barrier ($\Omega _{0}=0$, Figure \ref{angless}b) two
connected  fluxon-antivortex topological excitations appear to be
diametrically  opposed. In contrast to that, the greater is the angular
velocity $\Omega $  of the sheet beam, the closer to each other these two
groups of vortices are  located at the same moment of time, see Figures \ref%
{ang}b and \ref{bigang}b.

Finally, for the case of slowly rotating barrier ($\Omega _{1}=2\pi \times
0.11$ Hz) the antivortex, which is closest to the axis, moves towards the
central hole and annihilates with the central vortex (the red dashed~line),
as seen in Figure \ref{ang}c. Then, one of the vortices (the red line
nearest to the axis of the ring in Figure \ref{ang}d) finally drifts
towards  the central hole of the torus (Figure \ref{ang}e), while other
vortices and  antivortices move to the outer (low-density) region and
eventually decay  there. As the result, the vorticity of the~system tends to
be $m=+1$ (Figure  \ref{ang}d--f).


When angular velocity $\Omega $ in Equation~(\ref{rotation}) increases, the
symmetry of the trapping potential breaks even stronger, which can drive the
evolution of the system towards a \textit{nonrotating} ground state with $%
m=0  $ (zero vorticity), which is never observed not only for the
axially-symmetric potential, but~also in the case of weak symmetry breaking.
As one can see in Figure \ref{bigang}a--c, for $\Omega _{2}=2\pi \times
0.23$~Hz the~dynamics generally resembles that in Figure \ref{ang};
nevertheless,  the subsequent evolution leads to~the~nullification of the
final vorticity.  The point is that the vortex and antivortex which are
nearest to the inner  edge (Figure \ref{bigang}c) drift towards the center
and annihilate with  each other (Figure \ref{bigang}c--e). Other~vortices
and antivortices leave  the system, moving to the external periphery and
disappearing there (Figure  \ref{bigang}d--f).


\section{Conclusion}

\label{sec4}
We have investigated dynamics of quantum vortices associated with the
symmetry-breaking instability of superflows in interacting ring-shaped BECs.
We demonstrate that the tunneling across the~Bose Josephson junction in the
double-ring system is associated with the spontaneous breaking of the
rotational symmetry due to the formation of rotational fluxons (Josephson
vortices). We~demonstrate that dynamics of the merging contour-rotating
persistent currents is determined by the discrete rotational symmetry of the
system. We have studied in detail the relaxation process of the merging
counter-propagating persistent currents with vorticities $(+1,-1)$ and $
(+2,-2)$ in~the~framework of the weakly dissipative mean-field model. It
turns out that, when the population in~one of the rings slightly dominates
(i.e., the respective imbalance parameter does not reach a~threshold
(critical) value, $|P|<P_{\text{cr}}$), the final state of the system is
imposed by the \emph{less populated} ring. For instance, in the case of $
0<P<P_{\text{cr}}$, so that the population with vorticity $m_{1}=+1$
dominates over its counterpart with $m_{2}=-1$, the final state produced by
the merger has vorticity $m=-1$.
On~the~other hand,  it is found that the system with broken discrete
rotational symmetry and the~imbalance taking values below the critical
level, $|P|<P_{\text{cr}}$, can  be driven into a final state with any
vorticity, $m=+1,0,-1$, depending on  the symmetry-breaking perturbation of
the barrier potential which separates  the initial rings. For values of the
imbalance above the critical level, $ |P|>P_{\text{cr}}$, only two final
values of the vorticity ($m=+1$ or $0$  for $P>0$, and $m=-1$ or $0$ for $P<0
$) are observed as~a~result of the  merger.

These results may stimulate further investigation of the fundamental role of
the symmetry breaking in the evolution of quantum systems at the macroscopic
level in the ongoing experiments with matter-wave settings and atomtronic
circuits.
\section*{Acknowledgment}

The work of B.A.M. on this topic is supported, in part, by the Israel
Science Foundation through grant No. 1287/17. We wish to thank
O.G. Chelpanova for useful discussions.

\bibliographystyle{plain}
\bibliographystyle{unsrt}
\bibliography{Symmetry_Refs}

\end{document}